\documentclass[a4paper,11pt]{article}

\usepackage{pos}
\usepackage{lipsum} 
\usepackage{amsmath,amsthm,enumitem}
\usepackage{mathtools}
\usepackage{wrapfig}
\newcommand{\shrink}{\vspace{-0.3cm}}
\usepackage{lineno}

\usepackage{subcaption}

\title{Muon number reconstruction with IceTop using a two-component lateral distribution function}

\ShortTitle{Muon number reconstruction with IceTop using a two-component lateral distribution function}

\author{The IceCube Collaboration \\{\normalsize \normalfont(a complete list of authors can be found at the end of the proceedings)}\\}

\emailAdd{mark.weyrauch@kit.edu}

\abstract{

\noindent
The IceCube Neutrino Observatory, situated at the geographic South Pole, comprises both a surface component, IceTop, and a deep in-ice component. This unique setup allows for simultaneous measurements of low-energy ($\sim \rm{GeV}$) and high-energy ($\gtrsim 400\,\rm{GeV}$) muons generated in cosmic-ray air showers. The correlation between these low- and high-energy muons can serve as a valuable tool not only for analyzing cosmic-ray composition but also for tests of hadronic interaction models. 
However, IceTop does not feature dedicated muon detectors, making it challenging to measure the low-energy muon component for individual air showers.
\\
\noindent
For this reason, a two-component lateral distribution function is utilized for the simultaneous reconstruction of the primary energy and low-energy muon number on a single-event basis. This is achieved by combining analytical descriptions of the electromagnetic and muon lateral distributions. In this work, the underlying principles of this method will be discussed, as well as its capability for muon number reconstruction using the hadronic interaction models Sibyll 2.1, QGSJet-II.04, and EPOS-LHC. 

\vspace{4mm}

{\bfseries Corresponding authors:}
Mark Weyrauch$^{1*}$\\
{$^{1}$ \itshape Institute for Astroparticle Physics, Karlsruhe Institute of Technology}\\
$^*$ Presenter
}

\ConferenceLogo{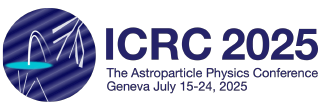}

\FullConference{39th International Cosmic Ray Conference (ICRC2025)\\
 15–24 July 2025\\
Geneva, Switzerland\\}

\begin{document}

\maketitle

\shrink
\section{Introduction}
\vspace{-0.1cm}
\noindent
Cosmic rays entering the Earth's atmosphere eventually interact with air molecules and produce a cascade of secondary particles, a so-called extensive air shower (EAS). EASs can be observed via ground-based detector arrays. However, because of the indirect nature of EASs measurements, the reconstruction of the characteristics of cosmic rays is subject to large systematic uncertainties. A significant contribution to the systematic uncertainty is constituted by the muon production in EASs, which varies significantly for different hadronic interaction models. Additionally, different experiments have reported a discrepancy between the number of muons measured in data as compared to the expectation from Monte-Carlo simulations \cite{Albrecht:2021cxw,Soldin:2021wyv,ArteagaVelazquez:2023fda}. Hence, tighter constraints on hadronic interaction models are required. In this context, the IceCube Neutrino Observatory allows the unique possibility to measure low-energy ($\sim$GeV) muons and high-energy ($\gtrsim 400\,\rm{GeV}$) muons for the same EASs. While the former is measured with the surface component of IceCube, IceTop, the latter can be observed by means of the deep in-ice array. In order to study correlations of low- and high-energy muons to test hadronic interaction models, a corresponding event-by-event based reconstruction is necessary. A single-event based high-energy muon reconstruction is possible using combined surface and in-ice information \cite{verpoest2023multiplicitytevmuonsair}. However, so far, the GeV muon density has only been measured by means of a statistical analysis in the $2.5-120\,\mathrm{PeV}$ energy range \cite{IceCubeCollaboration:2022tla}. In this work, the two-component lateral distribution function (LDF) is presented as a tool for the single-event based reconstruction of the low-energy muon number and primary energy. For this purpose, two LDFs are fit simultaneously to the charge distribution for EAS events observed with IceTop.

\shrink
\section{IceTop}
\vspace{-0.1cm}
\noindent
IceTop consists of 81 detector stations, each containing two ice-Cherenkov tanks deployed at a distance of $\sim 10 \,\rm{m}$ \cite{IceCube:2012nn}. The stations are distributed over an area of one square kilometer. The stations are spaced in a triangular grid with a distance of 125\,m, except for an infill area with a denser spacing of $\sim 40\,$m used for the measurement of low-energy EASs. Charged particles that traverse the ice volume inside the tanks generate Cherenkov light which is collected by digital optical modules (DOMs). Each tank is equipped with two DOMs for an increased dynamic range. If the charge deposit in one tank exceeds the discriminator threshold, this tank is classified as a `soft local coincidence' (SLC). In case both tanks in one station trigger within a time window of $1\,\rm{\mu}$s, this station has a `hard local coincidence' (HLC). The measured signal in each tank is calibrated to units of vertical equivalent muons (VEM), the average charge deposit produced by a vertically traversing muon. So far, the reconstruction of the energy and geometry of EASs with IceTop only includes HLCs \cite{IceCube:2012nn}. Since muons are more dominant far from the shower axis and therefore typically produce SLCs, both HLCs and SLCs are included in the two-component LDF reconstruction presented in this work.

\shrink
\section{Implementation of the two-component fit}
\vspace{-0.1cm}
\noindent
For the reconstruction of both the electromagnetic (EM) and muon contributions in air-shower events, two different lateral distributions are incorporated in a combined reconstruction procedure. The so-called `Double Logarithmic Parabola' (DLP) function \cite{DLPRef}
\begin{equation}
    S_{\rm{em}} = S_{\rm{em},125} \left( \frac{r}{r_{\rm{em}}} \right) ^{-\beta_{\rm{em}} - \kappa(S_{125}) \log_{10}{ \left( r/r_{\rm{em}} \right) }} , \ r_{\rm{em}} = 125\,\rm{m} \ , \label{eq:DLP}
\end{equation}
is used for the description of the EM ($e^\pm$,$\gamma$) part of the shower. The muon contribution is based on the Greisen LDF \cite{Greisen:1960wc}
\begin{equation}
    S_{\mu} = S_{\mu,550} \left(  \frac{r}{r_{\mu}} \right)^{-\beta_\mu} \left(  \frac{r+320\,\rm{m}}{r_{\mu}+320\,\rm{m}} \right)^{-\gamma(S_{125})} , \ r_\mu = 550\,\rm{m} \, ,
    \label{eq:Greisen} 
\end{equation}

\begin{figure*}[t!]
    \centering
    \begin{subfigure}[h!]{0.495\textwidth}
    \centering
    \includegraphics[width=\textwidth]{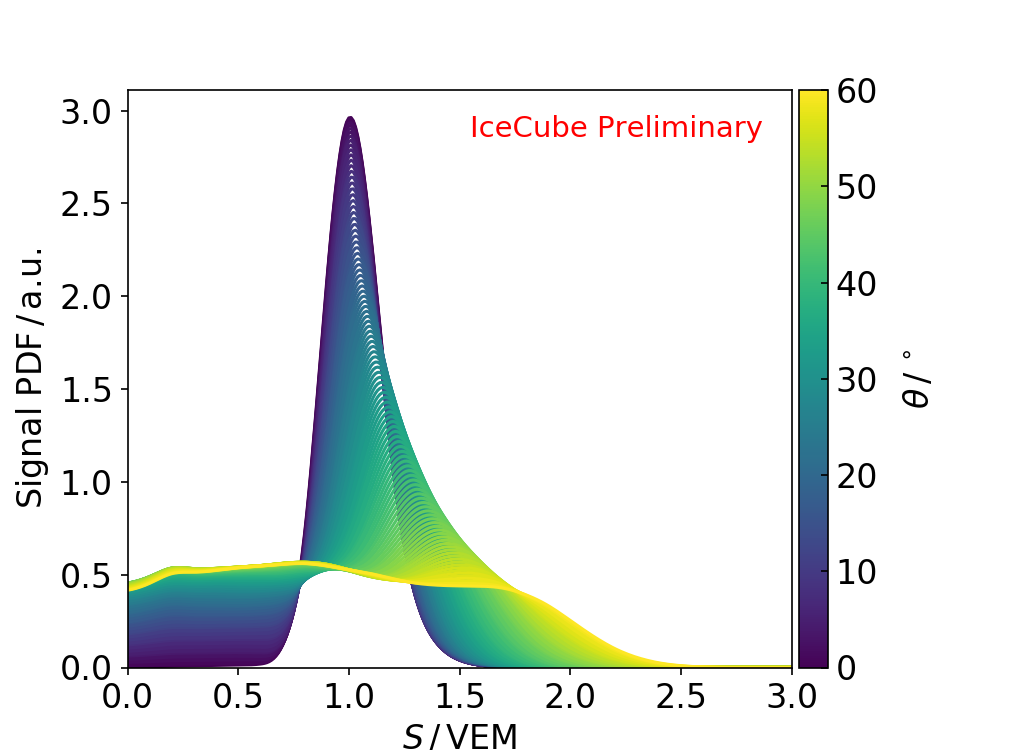}
    \end{subfigure}
    \hfill
    \begin{subfigure}[h!]{0.47\textwidth}
    \centering
    \vspace{0.6cm}
    \includegraphics[width=\textwidth]{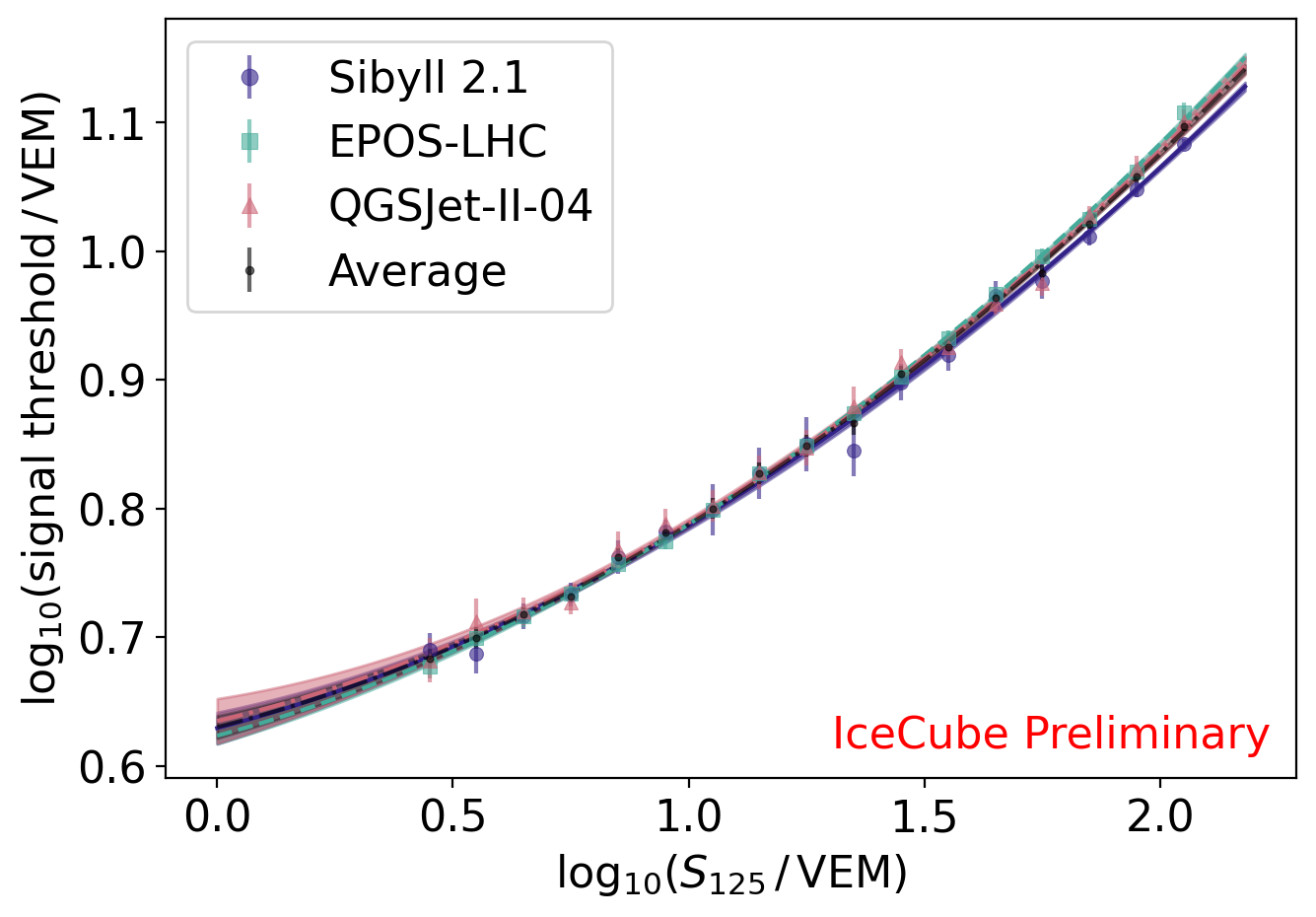}
    \end{subfigure}
    \shrink
    \caption{Left: Spline fits to the IceTop tank response for single muons. Right: Parametrization of the large signal threshold, $f_{\mu}(S_{125})$ \eqref{eq:p_signal}, used in the calculation of the HLC signal likelihood as a function of $\log_{10}S_{125}$.}
    \label{fig:n1_muon_splines}
    \shrink
\end{figure*}

\noindent
with the Greisen radius of $320\,$m. Both LDFs feature two slope parameters that are used to describe their shape. $\beta_{\rm{em}}$ and $\kappa$ are the slope and curvature of the DLP function, respectively. Only $\beta_{\rm{em}}$ is treated as a free parameter in the reconstruction. Analogously, for the muon LDF only $\beta_\mu$ is a fit parameter. $\kappa$ and $\gamma$ are both parametrized as functions of $S_{125}$, the energy proxy derived from the standard IceTop reconstruction (see \cite{IceCube:2012nn} for details), based on fits to average EM and muon lateral distributions.
The normalization of the LDFs is determined by $S_{\rm{em},125}$ and $S_{\mu,550}$, the signal strength at a given reference distance relative to the shower axis, $125\,$m and $550\,$m, respectively. While $S_{\rm{em},125}$ is used as a proxy for the primary energy, $S_{\mu,550}$ is used as a measure of the number of muons in the shower. For the reconstruction of the primary energy, a distance of $125\,\rm{m}$ combines a good reconstruction resolution with a small dependence on the primary type that initiates the air shower \cite{Andeen:2011}. Similar characteristics are found in the case of the muon number reconstruction using a reference distance of $550$\,m. The EM signal distribution can be described via a log-normal probability distribution functions (PDF) (see \cite{IceCube:2012nn} for details).
For the derivation of PDFs for the muon contribution, Geant4 \cite{GEANT4:2002zbu} simulations of the tank response for injected $E_\mu=1\,\rm{GeV}$ muons with different inclinations are applied for a given number of injected muons. The simulations are produced for a multiplicity of up to 15 muons. Above that, the PDF is approximated as a Gaussian. Fig. \ref{fig:n1_muon_splines} (left) shows the resulting PDF for one muon as a function of inclination and signal strength calibrated to units of VEM. The shape of the PDF is determined mainly by the geometry of the tank. The position of the peak\footnote{For an energy of $E_\mu=1\,\rm{GeV}$, the peak position for vertically injected muons is slightly above 1\,VEM.} shifts with $1/\cos{\theta}$ as the muon track length inside the sensitive volume increases. At the same time, the contribution of muons traversing the tank only on the edge (`corner clipping muons') increases, which leads to small signal deposits. These muon signal PDFs are saved as spline fits from which the probability $p_{\mu,\rm{sig}}(S|\theta, n)$  for a given inclination $\theta$, tank signal $S$, and muon multiplicity $n$, can be obtained during the reconstruction procedure. In this work, the muon inclination is assumed to follow the primary zenith angle. The signal PDFs are weighted with the Poisson probability to observe a given number of muons in order to obtain the total muon PDFs:
\begin{equation}
    p_\mu(S|\theta, \langle N_\mu \rangle) = \sum_{n}^{} \frac{\langle N_\mu \rangle^n}{n!} e^{-\langle N_\mu \rangle}  p_{\mu,\rm{sig}}(S|\theta,n) \ ,
    \label{eq:muPDF}
\end{equation}
with the expected number of muons, $\langle N_\mu \rangle$, as determined by the effective tank area ($A_{\rm{tank},\rm{eff}}=\pi r_{\rm{tank}}^2 \cos\theta + 2 h_{\rm{fill}} r_{\rm{tank}} \sin\theta$)\footnote{Since the signal per muon is proportional to its track length, the expected muon signal in VEM per tank can be converted to the expected muon number.} and the muon LDF ($\langle N_\mu \rangle = \langle S_\mu \rangle A_{\rm{tank},\rm{eff}}$). Here, $r_{\rm{tank}}$ and $h_{\rm{fill}}$ are the radius and fill height of the tank, respectively.
After deriving the PDFs for both the EM and the muon contribution, the combined likelihood can be constructed. For this purpose, the HLC and SLC signals are treated separately to account for the difference in the relative contribution of the EM and muon components to the signal deposit measured in full-station and single-tank hits:
\begin{gather}
\begin{aligned}
p_{\text{SLC}}\left( S|\theta , \langle S_{\rm{em}} \rangle, \langle S_{\mu} \rangle \right) \hspace{-2pt} &= \hspace{-2pt} \int_{0}^{S} \hspace{-8pt} p_{\rm{em}}(S_{\rm{em}}'|\theta,\langle S_{\rm{em}}\rangle c_{\rm{snow}}) p_{\mu}(S-S_{\rm{em}}'|\theta, \langle N_\mu \rangle) dS_{\rm{em}}' \, , \\
p_{\text{HLC}}\left( S|\theta , \langle S_{\rm{em}} \rangle, \langle S_{\mu} \rangle \right) \hspace{-2pt} &= \hspace{-2pt}
\begin{dcases}
     p_{\rm{em}}(S-\langle S_{\mu} \rangle|\theta,\langle S_{\rm{em}}\rangle c_{\rm{snow}}), & S > f_{\mu}(S_{125}) \, , \\
     p_{\mu}(0|\theta, \langle N_\mu \rangle) p_{\rm{em}}(S|\theta,\langle S_{\rm{em}}\rangle c_{\rm{snow}}) , & \log_{10}\left( S/\rm{VEM} \right) < -0.2 \, , \\[-4pt]
     \int_{0}^{S} p_{\rm{em}}(S_{\rm{em}}'|\theta,\langle S_{\rm{em}}\rangle c_{\rm{snow}}) \\[-8pt] \quad \quad p_{\mu}(S-S_{\rm{em}}'|\theta, \langle N_\mu \rangle) dS_{\rm{em}}' , & \text{else} \, .
  \end{dcases}
\end{aligned} \label{eq:p_signal}
\end{gather}
To account for the accumulation of snow on top of the IceTop tanks, the EM signal is modified by means of a snow correction factor \cite{IceCube:2012nn}, $c_{\rm{snow}}$. While for SLCs the likelihood calculation is based on a full convolution of the EM and muon PDFs, the HLC likelihood is divided into a small and large signal regime as well as an intermediate region to improve the reconstruction performance. 
In the small signal regime ($\log_{10} \langle S/\rm{VEM} \rangle < -0.2$), the average HLC signal is dominated by EM contribution (see \cite{WeyrauchECRS24} for details on the data driven derivation). In this region, the probability to measure a given signal strength is based on the EM probability. Since in Eq. \ref{eq:p_signal} the measured tank signal is used in the threshold condition, an additional weighting is applied with the probability of not observing any contribution of the muon signal, $p_{\mu}(0|\theta, \langle N_\mu \rangle)$. 
\begin{figure*}[t!]
    \centering
    \begin{subfigure}[h!]{0.48\textwidth}
    \centering
    \includegraphics[width=\textwidth]{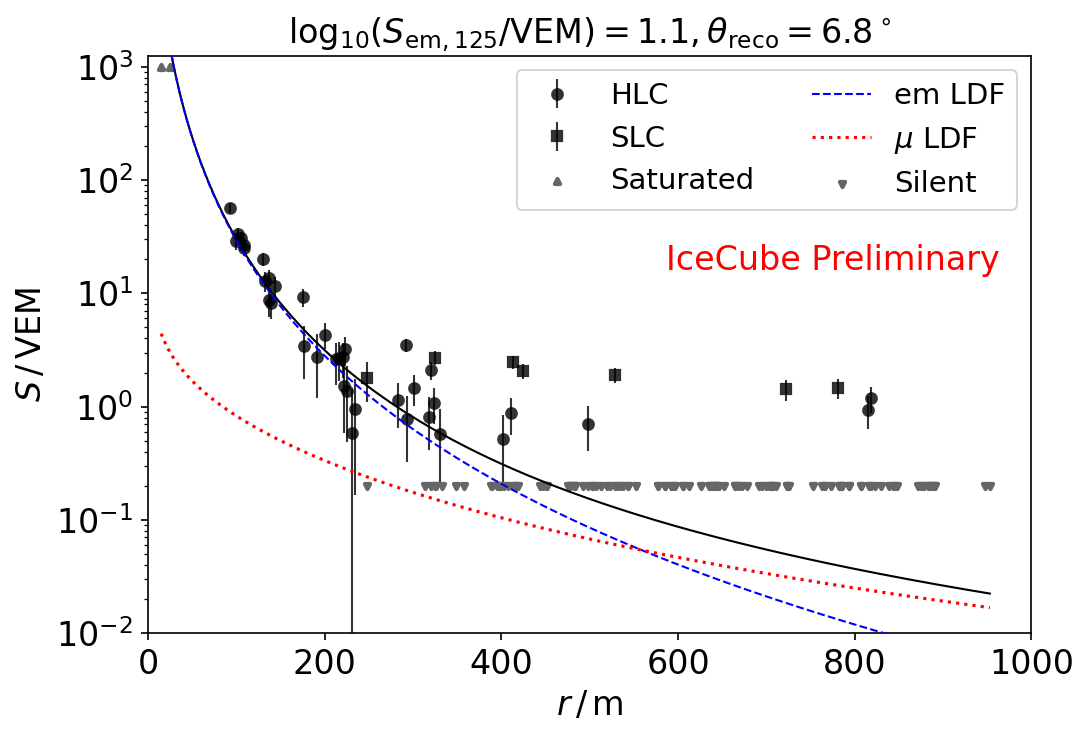}
    \end{subfigure}
    \hfill
    \begin{subfigure}[h!]{0.47\textwidth}
    \centering
    \includegraphics[width=\textwidth]{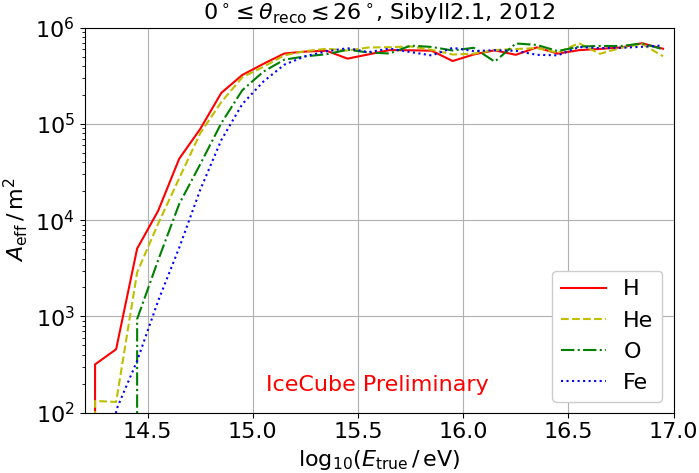}
    \end{subfigure}
    \shrink
    \caption{Left: Single event fit for a simulated EAS event ($\log_{10}(E_{\rm{true}}/\,\rm{eV})=16.16$). Silent detectors are drawn at a fixed value of $0.2\,$VEM. Right: Effective area for the two-component LDF reconstruction after application of all quality cuts for Sibyll 2.1.}
    \label{fig:single_event_fit}
\shrink
\end{figure*}
While the small signal regime is of importance for the region at large lateral distances, the large signal regime ($S > f_{\mu}(S_{125})$) is motivated by the region close to the shower axis. Since the hits in this region are vastly dominated by the EM component, only the EM probability is taken into account. The signal threshold for the large signal regime is chosen based on a MC study described in \cite{WeyrauchECRS24}. It allows the derivation of a shower-size dependent signal threshold above which the muon contribution to a measured HLC signal can be classified as non-significant. This procedure includes a particular fraction of phase space for the contribution of muons to the observed HLC signals. For this work, the inclusion of 85\% of the phase space is found to provide good reconstruction performance. The exact choice of threshold parametrization can be optimized for a given analysis goal. The parametrization, $f_{\mu}(S_{125})$, is shown in Fig. \ref{fig:n1_muon_splines} (right) for the hadronic interaction models Sibyll 2.1 (\cite{Sibyll2.1:2009}), EPOS-LHC (\cite{EPOS-LHC}) and QGSJet-II-04 (\cite{QGSJet}). Since the models show only small differences in the derivation of $f_{\mu}(S_{125})$, the parametrization based on the average of all three models is used. The combined log-likelihood can be written as 
\begin{gather}
\begin{aligned}
   \mathrm{llh} &= \log\left( p_{\text{HLC}}\right) + \log\left( p_{\text{SLC}}\right) + \mathrm{llh_{sat}} + \mathrm{llh_{sil,HLC}} + \mathrm{llh_{sil,SLC}} + \mathrm{llh_{t}} \\
   \rm{llh_{sil,HLC}} &= \log\left( 1 - p_{\rm{hit}}^2 \right) ; \rm{llh_{sil,SLC}} = \log\left( 1 - p_{\rm{hit}} \right) \ , \label{eq:tot_llh}
\end{aligned}
\shrink
\end{gather}
in which all single tank likelihoods are summed up to a global likelihood. The likelihood functions for saturated detectors, $\mathrm{llh_{sat}}$, and for the time fit, $\mathrm{llh_{t}}$, are defined analogously to the standard reconstruction (see \cite{IceCube:2012nn} for details). The likelihood of detectors without a trigger (`silent' detectors), $\rm{llh_{sil}}$, is based on the probability of not triggering a detector, $p_{\rm{nohit}}=1-p_{\rm{hit}}$, which is calculated separately for HLCs and SLCs by integrating the combined PDFs up to the threshold value, $s_{\rm{thr}}$
\begin{gather}
\begin{aligned}
p_{\rm{nohit},\rm{HLC}} &=
\begin{dcases}
\int_{S_{\rm{em}}=0}^{s_{\rm{thr}}} p_{\rm{em}}(S_{\rm{em}}'|\theta,\langle S_{\rm{tot}}\rangle) , \hspace{60pt} \log_{10} \langle S_{\rm{tot}}\rangle < -0.2 \, ,  \\
\hspace{-2pt} \sum_{n}^{} \hspace{-2pt} \frac{\langle N_\mu \rangle^n}{n!} e^{-\langle N_\mu \rangle} \hspace{-8pt} \int\displaylimits_{S_{\mu}=0}^{s_{\rm{thr}}} \hspace{-3pt} \int\displaylimits_{S_{\rm{em}}=0}^{s_{\rm{thr}}-S_{\mu}} \hspace{-14pt} p_{\mu,\rm{sig}}(S_{\mu}'|\theta,n)  p_{\rm{em}}(S_{\rm{em}}'|\theta,\hspace{-2pt} \langle S_{\rm{em}} \rangle c_{\rm{snow}}) , \text{else} \, . 
\end{dcases} \\
\end{aligned}
\end{gather}
The small signal regime is taken into account analogously to Eq. \ref{eq:p_signal} and $\langle S_{\rm{tot}}\rangle$ is the sum of the EM and muon signal expectation. 
For SLCs, $p_{\rm{nohit}}$ is always derived by integrating the convolved PDF. The reconstruction is performed as a three-step negative log-likelihood minimization (see \cite{Lesz:2023icrc} for an overview of the reconstruction framework). During reconstruction, the LDF parameters, $S_{\rm{em},125}$, $\beta_{\rm{em}}$, $S_{\mu,550}$ and $\beta_\mu$ as well as the shower geometry are successively varied (see \cite{WeyrauchECRS24} for details). Fig. \ref{fig:single_event_fit} (left) shows an example fit obtained after the last step. 

\section{Reconstruction performance}
\begin{figure*}[ht!]
    \centering
    \begin{subfigure}[h!]{0.49\textwidth}
    \centering
    \includegraphics[width=\textwidth]{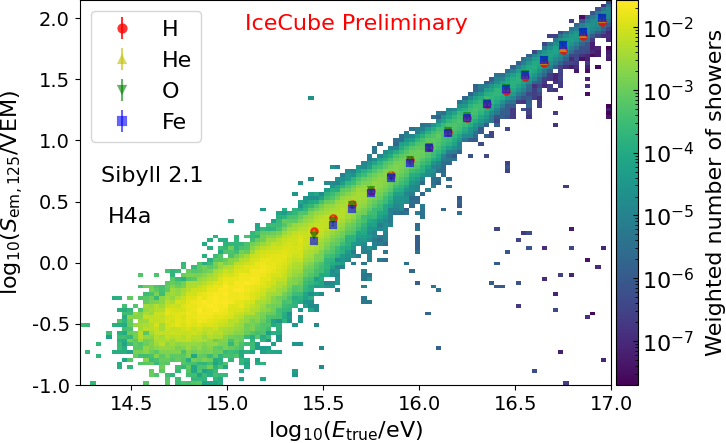}
    \end{subfigure}
    \hfill
    \begin{subfigure}[h!]{0.49\textwidth}
    \centering
    \includegraphics[width=\textwidth]{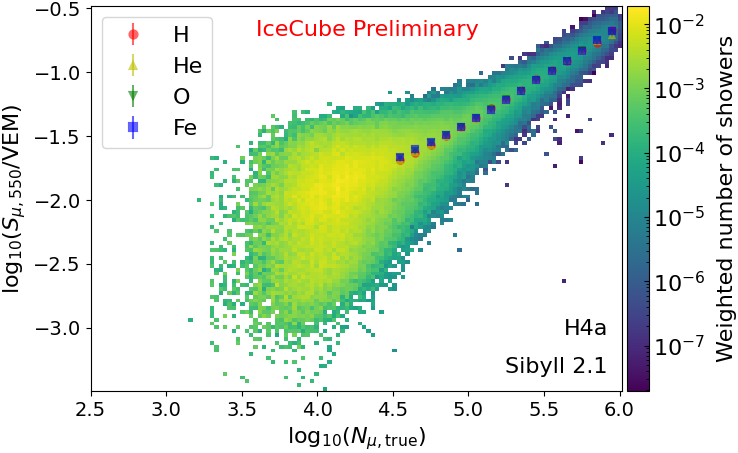}
    \end{subfigure}
    \shrink
    \caption{Energy proxy, $S_{\rm{em},125}$ (left), and muon number proxy, $S_{\mu,550}$ (right) as a function true energy and true muon number, respectively. Both distributions are weighted to H4a. The mean values are drawn approximately above the full efficiency threshold of the IceTop trigger and event selection.}
    \label{fig:Sibyll2.1_proxy_par}
\end{figure*}
\shrink 
\begin{figure*}[ht!]
\shrink
    \centering
    \begin{subfigure}[h!]{0.495\textwidth}
    \centering
    \includegraphics[width=\textwidth]{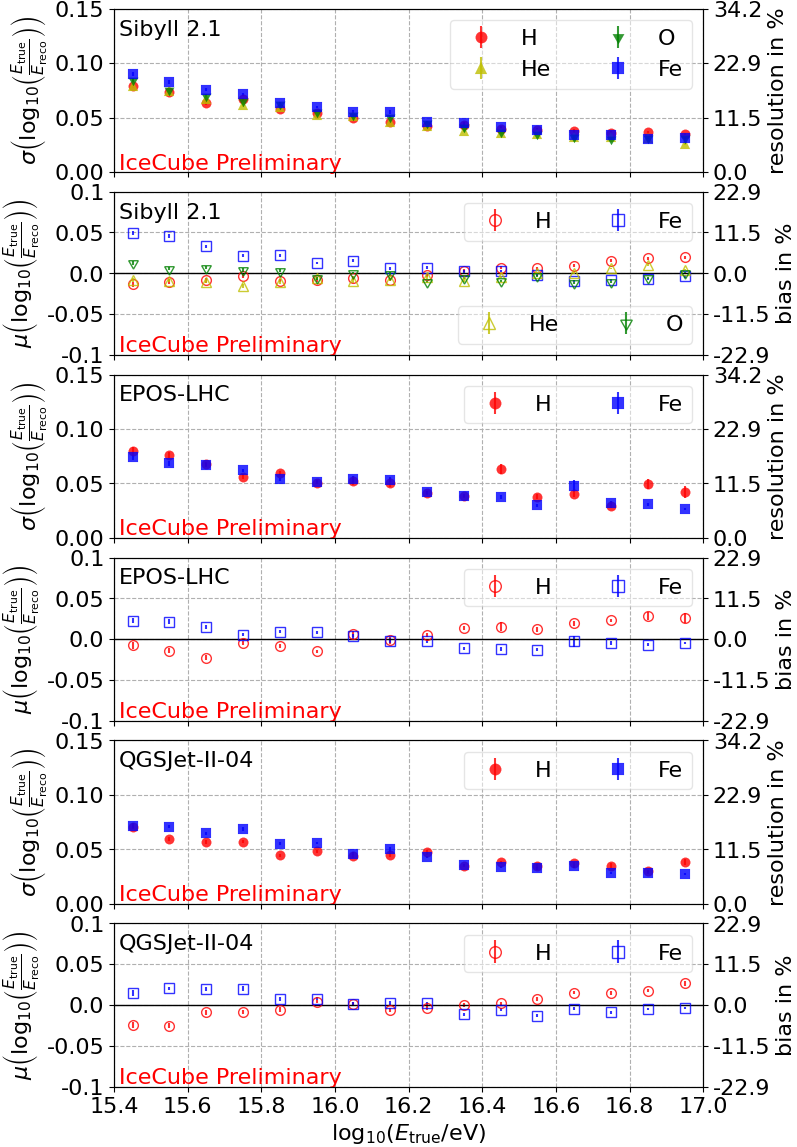}
    \end{subfigure}
    \hfill
    \begin{subfigure}[h!]{0.495\textwidth}
    \centering
    \includegraphics[width=\textwidth]{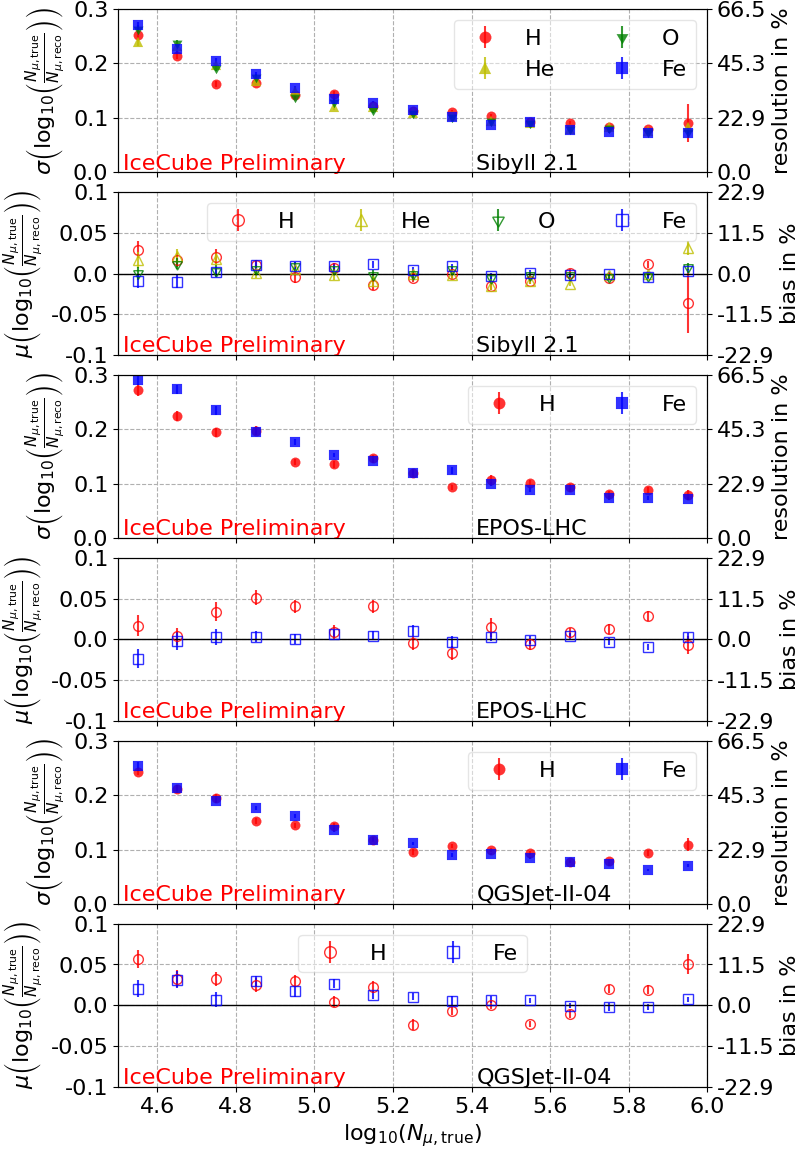}
    \end{subfigure}
    \shrink
    \caption{Resolution and bias for the energy (left column) and muon number (right column) reconstruction shown for Sibyll 2.1, EPOS-LHC and QGSJet-II-04 in the top, center and bottom row, respectively.}
    \label{fig:reconstruction_performance}
\shrink
\end{figure*}

\noindent
The reconstruction is applied to a simulation set for the snow overburden of the year 2012 produced with the hadronic interaction models Sibyll 2.1, EPOS-LHC and QGSJet-II-04, containing EASs with energies of $100\,\rm{TeV}-100\,\rm{PeV}$. While for Sibyll 2.1 EASs initiated by proton, helium, oxygen and iron are simulated, the simulation sets for the post-LHC models EPOS-LHC and QGSJet-II-04 only contain proton and iron primaries. 
Only near-vertical showers ($\theta \lesssim 26^\circ$) are included in the analysis. In addition, a set of quality cuts derived from previous IceTop analyses (see \cite{Aartsen:2013} for details) and $\log_{10} S_{\mu,550}>-3.5$ is applied. Moreover, a cut on the EM slope $\beta_{\rm{em}}$ as a function of the muon numer proxy $S_{\mu,550}$
\begin{equation}
    \beta_{\rm{em}} >
      \begin{cases}
         5\log_{10} S_{\mu,550} + 1 , & \log_{10} S_{\mu,550} \leq -1.27 \\
         2.2 , & \log_{10} S_{\mu,550} > -1.27
      \end{cases}
\end{equation}
\begin{wrapfigure}{R!}{0.45\textwidth}
 \centering
  \includegraphics[scale=0.38,angle=0]{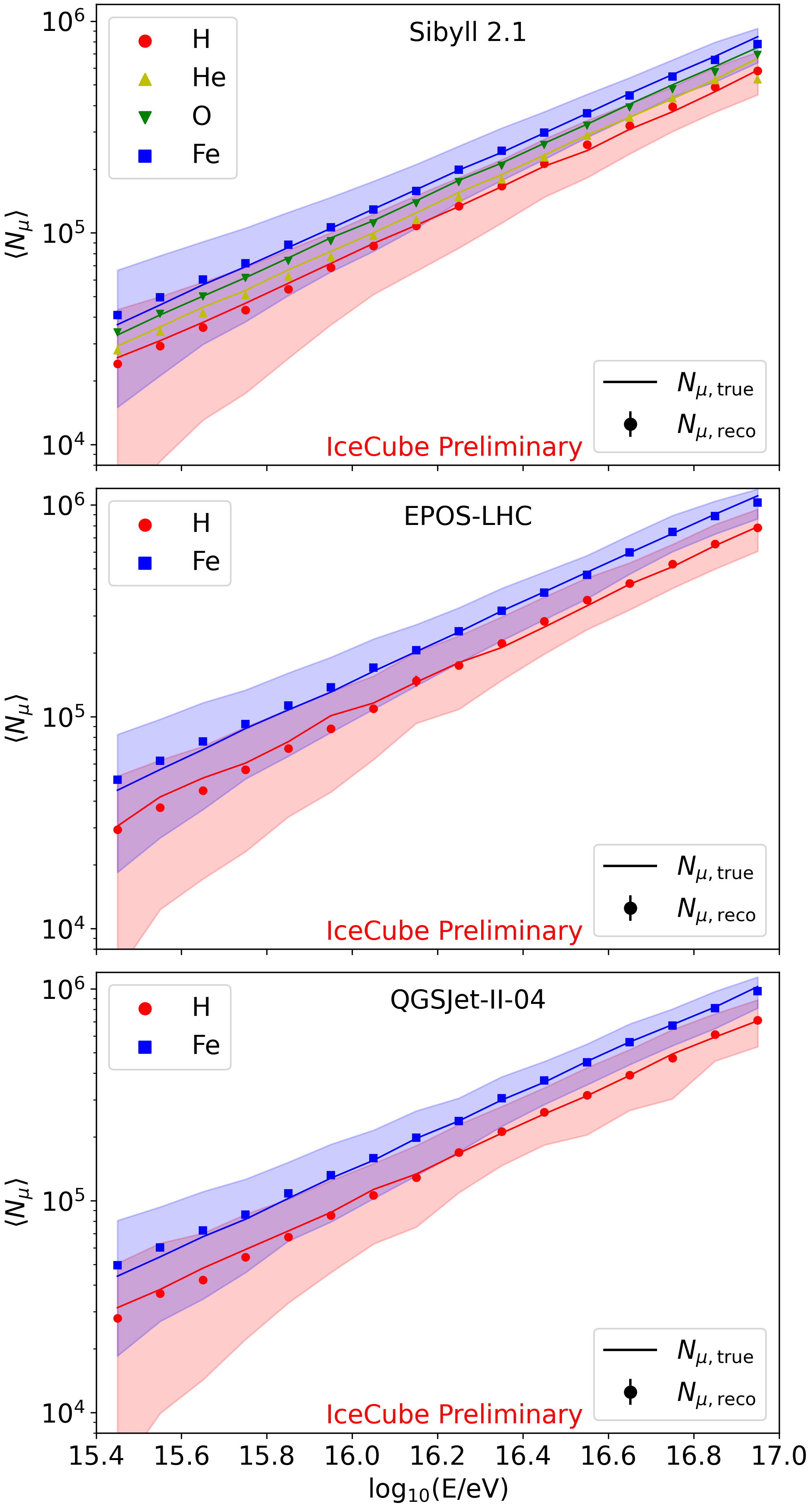}%
 \shrink \vspace{-0.05cm}
 \caption{Average muon number as a function of energy. Points and lines indicate the reconstructed and true quantities, respectively.}
 \label{fig:avg_muon_number_vs_energy}
 \vspace{-0.5cm}
\end{wrapfigure}

\noindent
removes remaining events with true core location outside of the IceTop array which are misreconstructed as contained events. 
Fig. \ref{fig:single_event_fit} (right) shows the effective area for the two-component LDF reconstruction after application of all quality cuts. The reconstruction reaches full efficiency for $\log_{10}(E/\,\rm{eV}) \gtrsim 15.4$ for all primaries, which is comparable to the standard reconstruction \cite{verpoest2023multiplicitytevmuonsair}. 
The resulting distributions of the energy proxy, $S_{\rm{em},125}$, and the muon number proxy, $S_{\mu,550}$, are shown in Fig. \ref{fig:Sibyll2.1_proxy_par} for Sibyll 2.1 as functions of the true energy and the muon number, respectively. The latter is determined for muons above 210\,MeV in energy and with lateral distances of less than 1\,km. The spread of these distributions is a measure of the reconstruction resolution, whereas a separation for different primary masses leads to a mass-dependent reconstruction bias. The distributions are weighted to the all-particle spectrum from the H4a flux model \cite{Gaisser:2011klf} and fitted with a calibration function to convert the proxy parameters to the corresponding reconstructed quantities. This procedure is applied separately for all three hadronic models. The resulting reconstruction resolution and bias are shown in Fig. \ref{fig:reconstruction_performance}. For all models, the primary energy can be reconstructed with only a small dependence on the primary mass on the few percent level. The muon number reconstruction does not show a significant mass dependence. For the given zenith range, the primary energy is reconstructed at the $\sim 12$\% level for energies above $\sim 10$\,PeV and improves to well below 10$\%$ toward $\sim 100$\,PeV for all models. The spread for the muon number reconstruction is larger compared to the energy reconstruction which translates to a resolution below 20$\%$ for showers around $100\,$PeV. Between the models, small differences are visible in the mass dependence and resolution of the reconstruction, which can be connected to differences, among others, in the position of the shower maximum, $X_{\rm{max}}$, the muon number, and the shape of the lateral particle distributions among the different models. 
Fig. \ref{fig:avg_muon_number_vs_energy} shows the average reconstructed muon number as a function of the reconstructed energy for all three models. To account for a remaining offset in the reconstructed points compared to the MC truth, the reconstructed average muon number was shifted by an energy dependent average correction factor based on an H4a composition assumption. The standard deviation for proton and iron primaries is indicated by a shaded area. The  primary energy and muon number can be well reconstructed for all models and for all primaries. Small offsets between the reconstructed points and the MC truth are caused mainly by the small mass dependence in the energy reconstruction.

\shrink
\section{Conclusion}
\vspace{-0.1cm}
\noindent
The two-component LDF combines two different model functions for the description of the electromagnetic and muon contribution of EAS events measured with IceTop. In this context, the signal expectations of the electro\-magnetic and muon LDF at a reference distance of 125\,m and 550\,m, respectively, can be utilized as a proxy for the primary energy and muon number. Using the proxy parameters, both quantities can be reconstructed simultaneously and on a single-event basis. For all studied hadronic interaction models, Sibyll 2.1, EPOS-LHC and QGSJet-II-04, the reconstruction of both primary energy and muon number is possible with minimal dependence on the primary mass (few percent level). For the energy reconstruction, a resolution of well below 10\% is achieved for showers of $\sim 100\,$PeV in energy. For the same showers, the muon number can be reconstructed with a resolution of below 20\%. Since the snow accumulation on top of the IceTop tanks increases throughout time, the electromagnetic signal is attenuated stronger for more recent datasets. Thus, the potential to isolate the muon contribution is increasing accordingly and can be studied in future analyses. In the context of tests of hadron\-ic in\-ter\-ac\-tion mod\-els and composition studies, the correlation of low- and high-energy muons is of high interest. The two-component LDF provides a tool for the single-event based low-energy muon reconstruction needed for such analyses. In addition, the two-component LDF slopes provide entirely new parameters for model tests \cite{Verpoest:2021_ICRC}, for which a preliminary study is presented in \cite{LincolnFahimDennisICRC25}. Another area of application are uncontained events, which can be used to significantly increase the statistics for EAS measurements with IceCube \cite{LillyICRC25}.

\bibliographystyle{ICRC}
\bibliography{ICRC2025_template_IceCube}

\clearpage

\section*{Full Author List: IceCube Collaboration}

\scriptsize
\noindent
R. Abbasi$^{16}$,
M. Ackermann$^{63}$,
J. Adams$^{17}$,
S. K. Agarwalla$^{39,\: {\rm a}}$,
J. A. Aguilar$^{10}$,
M. Ahlers$^{21}$,
J.M. Alameddine$^{22}$,
S. Ali$^{35}$,
N. M. Amin$^{43}$,
K. Andeen$^{41}$
% ...
\\
\\
% ...
$^{10}$ Universit{\'e} Libre de Bruxelles, Science Faculty CP230, B-1050 Brussels, Belgium \\
% ...
$^{16}$ Department of Physics, Loyola University Chicago, Chicago, IL 60660, USA \\
$^{17}$ Dept. of Physics and Astronomy, University of Canterbury, Private Bag 4800, Christchurch, New Zealand \\
% ...
$^{21}$ Niels Bohr Institute, University of Copenhagen, DK-2100 Copenhagen, Denmark \\
$^{22}$ Dept. of Physics, TU Dortmund University, D-44221 Dortmund, Germany \\
% ...
$^{35}$ Dept. of Physics and Astronomy, University of Kansas, Lawrence, KS 66045, USA \\
% ...
$^{39}$ Dept. of Physics and Wisconsin IceCube Particle Astrophysics Center, University of Wisconsin{\textemdash}Madison, Madison, WI 53706, USA \\
% ...
$^{41}$ Department of Physics, Marquette University, Milwaukee, WI 53201, USA \\
% ...
$^{43}$ Bartol Research Institute and Dept. of Physics and Astronomy, University of Delaware, Newark, DE 19716, USA \\
% ...
$^{63}$ Deutsches Elektronen-Synchrotron DESY, Platanenallee 6, D-15738 Zeuthen, Germany \\
$^{\rm a}$ also at Institute of Physics, Sachivalaya Marg, Sainik School Post, Bhubaneswar 751005, India \\
% ...

\subsection*{Acknowledgments}

\noindent
The authors gratefully acknowledge the support from the following agencies and institutions:
USA {\textendash} U.S. National Science Foundation-Office of Polar Programs,
U.S. National Science Foundation-Physics Division,
U.S. National Science Foundation-EPSCoR,
U.S. National Science Foundation-Office of Advanced Cyberinfrastructure,
Wisconsin Alumni Research Foundation,
Center for High Throughput Computing (CHTC) at the University of Wisconsin{\textendash}Madison,
Open Science Grid (OSG),
Partnership to Advance Throughput Computing (PATh),
Advanced Cyberinfrastructure Coordination Ecosystem: Services {\&} Support (ACCESS),
Frontera and Ranch computing project at the Texas Advanced Computing Center,
U.S. Department of Energy-National Energy Research Scientific Computing Center,
Particle astrophysics research computing center at the University of Maryland,
Institute for Cyber-Enabled Research at Michigan State University,
Astroparticle physics computational facility at Marquette University,
NVIDIA Corporation,
and Google Cloud Platform;
Belgium {\textendash} Funds for Scientific Research (FRS-FNRS and FWO),
FWO Odysseus and Big Science programmes,
and Belgian Federal Science Policy Office (Belspo);
Germany {\textendash} Bundesministerium f{\"u}r Forschung, Technologie und Raumfahrt (BMFTR),
Deutsche Forschungsgemeinschaft (DFG),
Helmholtz Alliance for Astroparticle Physics (HAP),
Initiative and Networking Fund of the Helmholtz Association,
Deutsches Elektronen Synchrotron (DESY),
and High Performance Computing cluster of the RWTH Aachen;
Sweden {\textendash} Swedish Research Council,
Swedish Polar Research Secretariat,
Swedish National Infrastructure for Computing (SNIC),
and Knut and Alice Wallenberg Foundation;
European Union {\textendash} EGI Advanced Computing for research;
Australia {\textendash} Australian Research Council;
Canada {\textendash} Natural Sciences and Engineering Research Council of Canada,
Calcul Qu{\'e}bec, Compute Ontario, Canada Foundation for Innovation, WestGrid, and Digital Research Alliance of Canada;
Denmark {\textendash} Villum Fonden, Carlsberg Foundation, and European Commission;
New Zealand {\textendash} Marsden Fund;
Japan {\textendash} Japan Society for Promotion of Science (JSPS)
and Institute for Global Prominent Research (IGPR) of Chiba University;
Korea {\textendash} National Research Foundation of Korea (NRF);
Switzerland {\textendash} Swiss National Science Foundation (SNSF).

\end{document}